\documentclass[12pt,preprint]{aastex}

\begin{document}

\title{Multi-Epoch Observations of the Redwing Excess in the Spectrum of 3C279}
\author{Brian Punsly\altaffilmark{1}} \altaffiltext{1}{1415 Granvia Altamira, Palos Verdes Estates CA,
USA 90274 and ICRANet, Piazza della Repubblica 10 Pescara 65100,
Italy, brian.punsly@verizon.net or brian.punsly@comdev-usa.com}

\begin{abstract}
It has been previously determined that there is a highly significant
correlation between the spectral index from 10 GHz to 1350 $\AA$ and
the amount of excess luminosity in the red wing of quasar CIV $\lambda$1549 broad
emission lines (BELs). Ostensibly, the prominence of the red
excess is associated with the radio jet emission mechanism and is
most pronounced for lines of sight close to the jet axis. Studying
the scant significant differences in the UV spectra of radio loud
and radio quiet quasars might provide vital clues to the origin of
the unknown process that creates powerful relativistic jets that
appear in only about ten percent of quasars. In this study, the
phenomenon is explored with multi-epoch observations of the MgII $\lambda$2798
broad line in 3C 279 which has one of the largest known redwing
excesses in a quasar spectrum. The amount of excess that is detected appears to be independent
of all directly observed optical continuum, radio or submm
properties (fluxes or polarizations). The only trend that occurs
in this sparse data is: the stronger the BEL, the larger the fraction of flux
that resides in the redwing. It is concluded that more monitoring is
needed and spectropolarimetry with a large telescope is essential
during low states to understand more.
\end{abstract}


\section{Introduction}Perhaps the greatest mystery of the quasar
phenomenon is that $\sim 10\%$ of the quasar population possess
powerful relativistic radio jets (known generically as radio loud
quasars RLQs). These are dramatic features that can transport
$\gtrsim 10^{40}$ W hundreds of kiloparsecs into the intra-cluster
medium \citep{wil99}. Yet, the majority of quasars are radio quiet quasars
(RQQs) that are defined by "weak" jet power. In spite of this most
conspicuous property, the optical/UV continuum and broad emission
lines (BELs) in RLQS and RQQs are remarkably similar
\citep{cor94,zhe97,tel02}. Systematic differences in the two
families of spectra are only revealed by the study of subtle lower
order spectral features \citep{cor98,ric02}. Perhaps the most
extreme difference in the spectra of these two classes of objects is
a highly significant correlation between the spectral index from 10
GHz to 1350 $\AA$ and the amount of excess luminosity in the red
wing of quasar CIV $\lambda$1549 broad emission lines (BELs), at $>99.9999\%$
statistical significance \citep{pun10}. The prominence of the
redward excess is apparently associated with the radio jet emission
mechanism and is most pronounced for lines of sight close to the jet
axis. \footnote{It should be noted that a subtlety of the correlation
is that some lobe dominated quasars have significant asymmetry. The
most extreme case in Figure 1 of \citet{pun10} is PKS 0454-020 with
an asymmetry larger than most blazars.} In some blazars (core
dominated radio sources), the asymmetry is so extreme that the line
shape becomes similar to a right triangular, the emission is
dominated by the red gently sloping side ("the hypotenuse"). In this
paper, we explore why this effect is so pronounced in some radio
loud AGN (active galactic nuclei) that are viewed along the jet
axis.

\par The red excess in the BELs has been noted in anecdotal examples of blazar spectra
including the first case to be studied in detail, H$\alpha$ in 4C
34.47 in \citet{cor97}. The origin of the redwing excess is unknown
and explanations have ranged from gravitational and transverse
redshift within 200 M (gravitational radii) from the central black
hole, reflection off optically thick, out-flowing clouds on
the far side of the accretion disk or transmission through inflowing
gas on the near side of the accretion disk
\citep{cor97,net95,mar96}. None of these scenarios are
substantiated, but are speculations that are designed to describe a difficult to understand
observation. The archives of observational data seem inadequate for
determining a likely explanation.
\par In \citet{net95}, HST data were used to show that 3C 279 had very asymmetric CIV $\lambda$1549 and MgII $\lambda$2798 BELs.
In this study, we explore the variation of the MgII $\lambda$2798 line profile
over time in 3C 279. This is an interesting case study because 3C
279 is perhaps the most conspicuous blazar in the local Universe
(z=0.5356 from \citet{net95}). It has the most extreme blazar properties, optical
polarization larger than 20\%, extreme variability in all
observable bands, superluminal jet speeds exceeding 20c and it is
one of the brightest gamma ray sources
\citep{abd09,ghi10,imp90,bon12,lar12,lis09,jor11}. 3C 279 also has a
spectrum with some of the most redward asymmetric BELs ever
observed. As a consequence of the extraordinary nature of this blazar, there is a
wealth of simultaneous long term monitoring data from radio waves to
gamma rays which facilitates multi-epoch comparisons. Since the
effect is most pronounced in blazars, one is often forced to deal
with the huge amount of dilution from the optical, high frequency
tail of the Doppler enhanced synchrotron emission from the
relativistic jet \citep{lin85}. Thus, in 3C 279 for example, it is difficult to
extricate the BELs from the continuum unless the jet emission is
weak (a low state).
\begin{figure}
\includegraphics[width=160 mm, angle= 0]{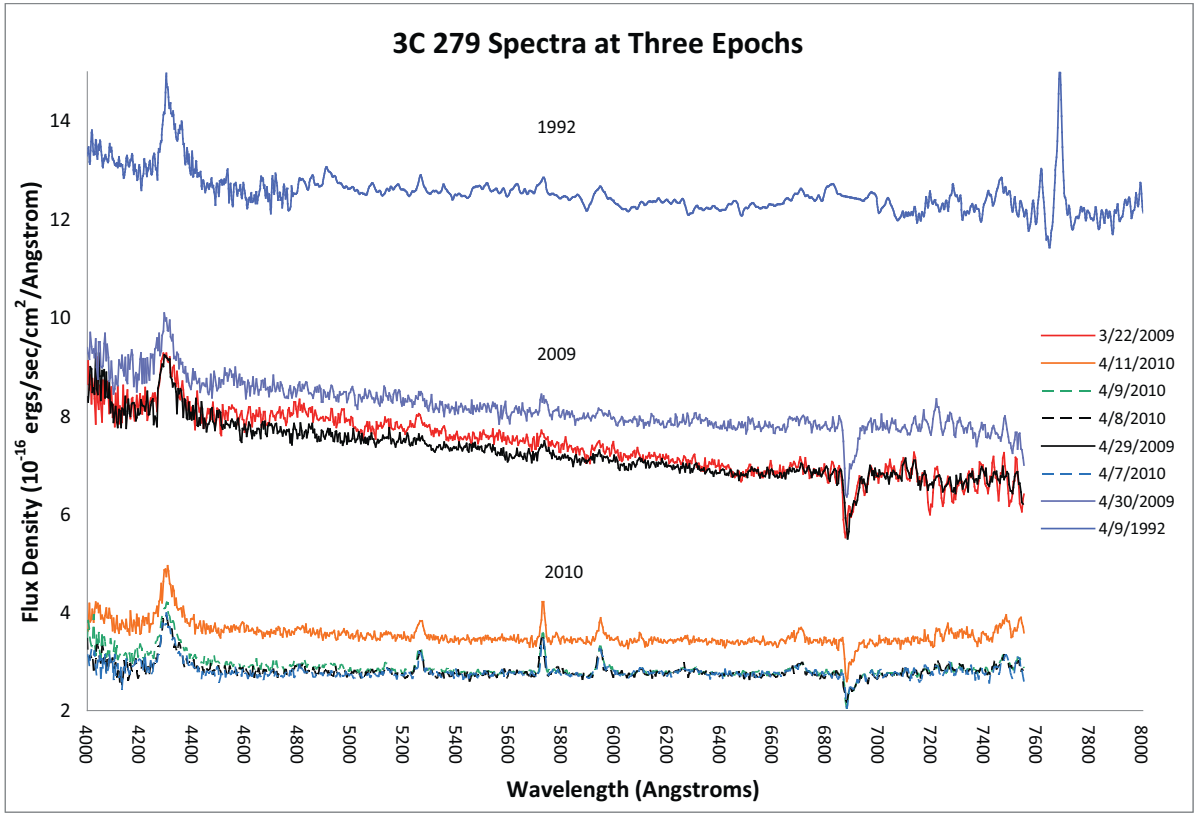}
\caption{Optical spectra at three different epochs.}
\end{figure}
\begin{figure}
\includegraphics[width=160 mm, angle= 0]{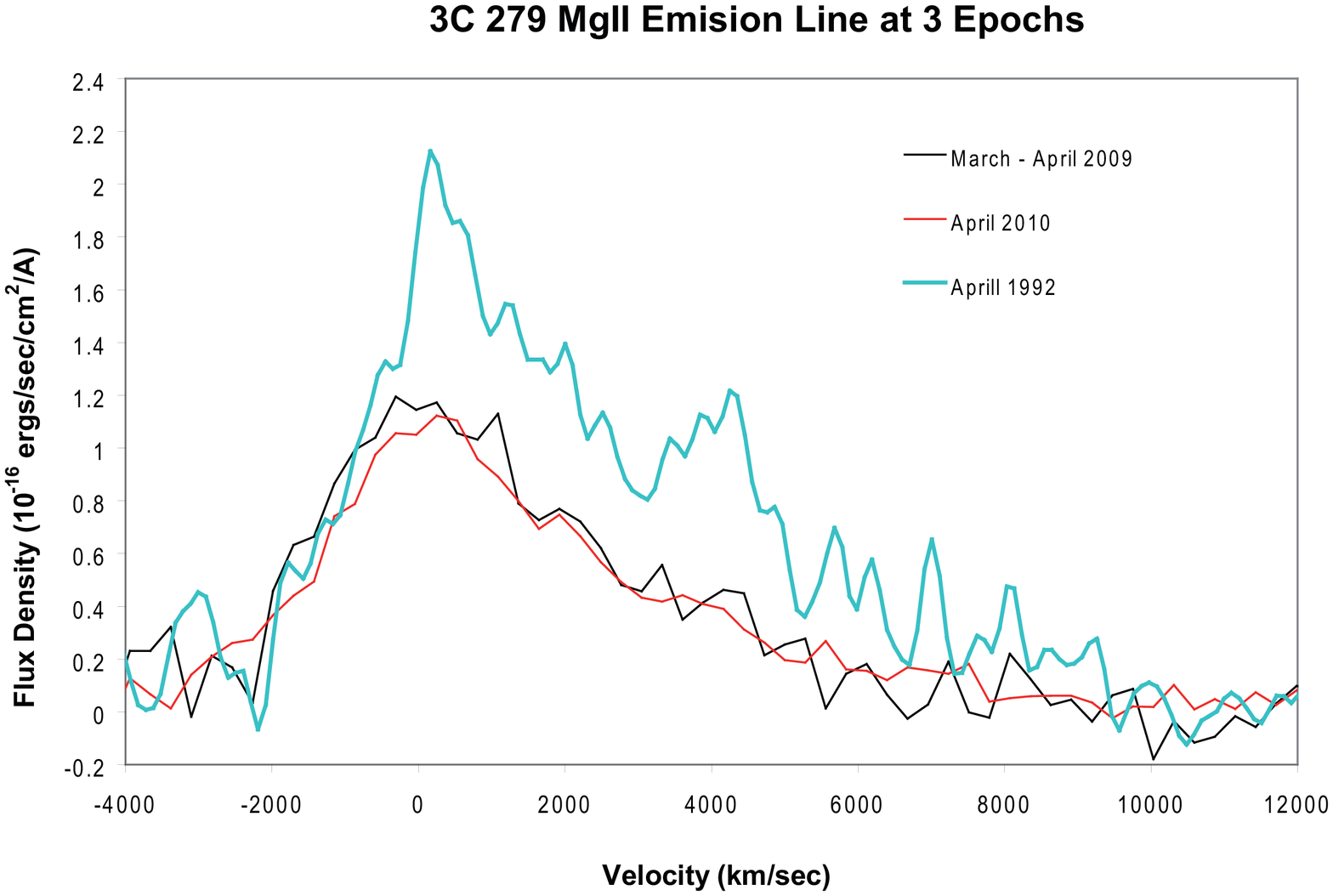}
\caption{Comparison of MgII $\lambda$2798 profiles in velocity space at three
different epochs that are spread over 18 years.}
\end{figure}
\section{Spectra From Three Epochs} The "Ground-based Observational Support of the Fermi Gamma-ray Space Telescope
at the University of Arizona," \citet{smi09} provides an opportunity
to find high signal to noise MgII $\lambda$2798 spectra of 3C 279. The required combination of a weak
synchrotron component and good seeing is not common, but with
long term monitoring, the chance of a high signal to noise observation of the MgII $\lambda$2798 BEL is vastly
improved. The frequent monitoring with Steward Observatory also
allows for the binning of data, thereby improving the signal to
noise ratio in the redwing of the MgII $\lambda$2798 BEL. Figure 1 plots the
optical spectra at three distinct epochs, covering 18 years.
\par The first epoch is April 9, 1992. The blue frequencies are from the HST, Faint Object
Spectrograph with the G400 grating. The longer wavelengths are from
the 2.7 m telescope at the McDonald Observatory using the Large Cassegrain
Spectrograph. These data were previously discussed in \citet{net95}.
The fully reduced and calibrated data were generously provided by B.
Wills. The raw spectral data were tabulated every $\approx$ 0.35 \AA\
and then binned and smoothed to a resolution of $\approx$ 1.4 \AA\ .
The details of the data processing can be found in
\citet{net95,wil95}. The continuum during April 1992 was the brightest of the
three epochs. Fortunately, the MgII $\lambda$2798 BEL was very strong at this
time. Other features to note are a similarly asymmetric CIV $\lambda1549$ BEL from
the same observing run with the G270 grating (not plotted, but see
\citet{net95}) and a strong [OIII] $\lambda$5007 narrow line.
\par The most recent epoch is the faintest, spanning the interval from April 7, 2010 to April 11, 2010.  The data are from the 3m Bok Telescope
of the Steward Observatory \citep{smi09}. The highest signal to
noise epochs from this time interval were collected (and plotted in
Figure 1) for future binning. The continuum in this state is an
order of magnitude fainter than in 1992, and the asymmetric broad
wing is clearly visible. Another interesting feature to note is the
distinct [OII] $\lambda$3727 narrow line. A second epoch from the Steward
Observatory archive is a year earlier with a continuum luminosity
intermediate between the other two epochs.
\par Figure 2 is a plot of the region of the spectrum that is contiguous to the Mg II BEL in velocity space. The line
profiles are obtained after subtraction of a linear fit to the
continuum from 4050 \AA\ to 4750 \AA\ . The HST data from B. Wills
are not the data used to plot the Mg II line profile that was
displayed in velocity space in Figure 2 of \citet{net95}. That low
resolution spectrum ($\sim 4 \AA$) was from a 15 minute observation
with the 4m CTIO Telescope using the RC spectrograph 4 days later.
The Steward Observatory line profiles were binned as a weighted average with a weight that
is inversely proportional to the RMS noise in the adjacent
continuum. Figure 2 indicates that the Mg II broad-line is virtually
unchanged (within the accuracy of the observations) from 2009 to
2010, but drastically weaker than in 1992. This would seem to
indicate a time scale for changes in the red wing excess that is
larger than 1 year and less than 18 years. This is in
contradistinction to jet emission that has order of unity variations
on the orders of hours to weeks \citep{lar12}. Thus, one does not expect to find trends
between the amount of red excess and the rapid observed flux variations from the Doppler enhanced jet.
\section{Characterizing the Red Asymmetry} In order to characterize the time
variability in the MgII $\lambda$2798 BEL depicted in Figure 2 requires a method of quantifying
the redward line asymmetry. The amount of perceived red excess is not uniquely defined and three
nonparametric methods for describing its magnitude are considered in Table 1.
\par The study in \citet{wil95} that included 3C 279 quantified the red asymmetry by $A_{25-80}$.
The quantity, $A_{25-80}$, is defined in terms of the full width
half maximum, FWHM, in $\AA$, the midpoint of an imaginary line
connecting a point defined at 1/4 of the peak flux density of the
BEL on the red side of the BEL to 1/4 of the peak flux density on
the blue side of the BEL, $\lambda_{25}$, and a similar midpoint
defined at 8/10 of the flux density maximum,  $\lambda_{80}$, as
\begin{equation}
A_{25-80} = \frac{\lambda_{25} - \lambda_{80}}{FWHM} \;.
\end{equation}
A positive value of $A_{25-80}$ means that there is excess flux in
the red broad wing of the BEL. A negative value of $A_{25-80}$ indicates a blueward
asymmetry of the BEL. In order to calculate $A_{25-80}$ in the presence of the
random noise that is superimposed on the line profiles in Figure 2, one can proceed
as in \citet{wil95}. Using the continuum fit described in the last section,
the lines are fit by two or three Gaussians profiles which interpolates between the
fluctuations of the random noise. The values for each epoch are listed in column (5) of Table 1.
\par The errors associated with estimating $A_{25-80}$ arise
primarily from the uncertainty in $\lambda_{25}$
since the signal to noise ratio is the smallest in the broad wings.
The error in each quantity in equation (1) was individually
estimated and the results were added in quadrature. The error in
$\lambda_{25}$, for example, was achieved by approximating the
region near the 1/4 maximum point of the line profile by the composite
Gaussian fit. The error in $\lambda_{25}$ was the determined to be
slope of this composite Gaussian fit ($\partial \lambda / \partial F_{\lambda}$)
at the 1/4 maximum point times the RMS noise level. This naturally
produces larger errors in $A_{25-80}$ for epochs with very broad
wings, i.e., the more horizontal the spectrum in the wings, the
larger the slope ($\partial \lambda / \partial F_{\lambda}$) will
be. These uncertainties are listed in column (5) of Table 1.
\par The line luminosities in Table 1 are computed using the
Galactic extinction from \citet{sch11} as posted in the NASA
Extragalactic Database and a cosmological model defined by
$H_{0}$=70 km/s/Mpc, $\Omega_{\Lambda}=0.7$ and $\Omega_{m}=0.3$.
The first thing to notice is that the 2009 and 2010 line profiles
are virtually identical in Figure 2 and Table 1 even though the
optical polarization is an order of magnitude larger in 2009. The line was
decomposed into two empirical components, a blue side (with negative velocity in Figure 2)
and a red side based on a redshift, z = 0.5356. Column (2) of Table 1 is the total line luminosity, $L(\mathrm{Mg II})$,
and column (3) is the ratio of red side luminosity, $L(\mathrm{Mg II})_{\mathrm{red}}$, to the
blue side luminosity, $L(\mathrm{Mg II})_{\mathrm{blue}}$. Column (4) is the red excess
that is defined by reflecting the blue side about the zero velocity axis and computing the red residuals.
This quantity is then normalized by $L(\mathrm{Mg II})$,
\begin{equation}
\mathrm{Red \; Excess} \equiv \frac{L(\mathrm{Mg II}) - 2L(\mathrm{Mg II})_{\mathrm{blue}}}{L(\mathrm{Mg II})}\,.
\end{equation}
The next column is the asymmetry parameter, $A_{25-80}$, defined in Equation (1). The last two columns depict
the strength and polarization of the optical continuum.
\begin{table}
\caption{Mg II Emission Line and Optical Continuum Properties}
{\footnotesize\begin{tabular}{cccccccc} \tableline\rule{0mm}{3mm}
Epoch & $L(\mathrm{Mg II})$ & $\frac{L(\mathrm{Mg II})_{\mathrm{red}}}{L(\mathrm{Mg II})_{\mathrm{blue}}}$  & Red  & $A_{25-80}$ & 6500 \AA\ Flux Density & Continuum  \\
 & ($10^{43}$ erg/s) & & Excess & & (ergs/ $\mathrm{cm}^{2}$ /s/ \AA\ ) & Polarization \\
\tableline \rule{0mm}{3mm}
1992 & $1.78 \pm 0.31 $ & $2.34\pm 0.58$ & $0.40\pm 0.07$  & $0.55 \pm 0.12$ & $1.32 \times 10^{-15}$  &   ...  \\
2009 & $0.96 \pm 0.28 $  & $1.56 \pm 0.48$  & $0.22 \pm 0.05$ & $ 0.24 \pm 0.13$ & $8.44 \times 10^{-16}$ &  10.1 \% - 21.1 \%     \\
2010 & $0.96 \pm 0.15 $  & $1.47 \pm 0.26$ & $0.19 \pm 0.03$ & $0.27 \pm 0.07$ & $2.92 \times 10^{-16}$  & 1.4 \% - 2.9 \%  \\
\end{tabular}}
\end{table}
\par The errors in column (2) of Table 1 represent the luminosity of the RMS of the residuals (noise level) to the composite
Gaussian fit that is obtained by integrating the residual luminosity over the entire line profile
from -5,000 km/sec (blue) to 13,000 km/sec (red).  Similarly, one computes the uncertainty in the blue (red) side
of the line decomposition as the integrated luminosity of the RMS residual noise level (note that the RMS residual noise level
is always positive, by definition, even when the residual luminosity is neagtive) over the blue (red) portion of the line profile
from -5,000 km/sec to 0 km/sec (0 km/sec to 13,000 km/sec). The errors in the blue and red sides
propagate through quadratures to generate the error estimates in columns (3) and (4).
\par The intent of presenting Table 1 is to demonstrate that the variation in the red wing excess over
an 18 year time span exists independent of how it is defined. The last two rows show that
a virtually unchanged line profile can still be clearly detected even when the synchrotron background triples in strength.
Thus, the Steward Observatory monitoring can detect a well-defined asymmetric profile over a wide range of continuum luminosity
and should provide a wealth of epochs with asymmetric profiles in the coming years. The preliminary trend in the data is
that when the line strength is elevated, the degree of asymmetry increases.

\section{Discussion}The purpose of this Letter was not to make
strong claims as to the physical origin the of red wing excess in
blazars. The purpose is to show that it is possible to see changes
in the red wing excess in blazars.
\par The primary result of Table 1 is that the MgII redwing excess does not vary
in consort with the rest of the broad line emission. Considering the conspicuous
jet viewed in a nearly pole-on orientation, it seems likely that the red excess
is associated with the jet and not the virialized gas responsible for the
core and blue wing of the emission line. This seems to favor the optically thick
outflow scenario described in the introduction. The phenomenon might be related to the outflow
that is responsible for the broad absorption lines seen in some polar orientation
broad absorption line quasars \citep{zho06,gho07,pun11}.
\par Temporal variations in line properties can be the compared
to contemporaneous variations of jet properties on subparsec scales that can be detected
during the extensive monitoring from milimeter wavelengths to gamma rays of many FERMI
selected blazars and high frequency (43 GHz) VLBA images \citep{ata11,bon12,jor11,lar12}. In principle,
any such causal connections can be be used as a probe of possible relationships between jet propagation/formation and
the gaseous environment on subparsec scales, the broad line region.
This comment in one way attempts to minimize the complexities introduced by large
Doppler enhancement of the emission that can make it extremely difficult
to determine intrinsic properties of the jet.

\par The data presented here from three epochs observed with modest
apertures are insufficient to properly explore the possibilities
noted above. To improve our understanding requires more optical data
with larger telescopes and long term monitoring. Future long term
monitoring (since the time scale for variation is longer than one
year from Table 1) of the FWHM and the luminosity using the
decompositions into red, blue and core components can be used as a
crude surrogate for reverberation mapping of the red wing (since the
optical continuum is hidden by the synchrotron emission). It would
be important to find time lags (if any) between the blue wing, the line core and the red excess.
This would yield information on the
relative locations of the gas producing the line core and the gas responsible
for the red excess. Perhaps of more interest, it is proposed that the Steward
Observatory monitoring be used as a trigger for a large telescope
observation. If a broad line is clearly displayed in the Steward
Observatory data this would trigger a large (8m) telescope to
observe 3C 279 with high resolution spectropolarimetry. The purpose
being the same as in the study of broad absorption line quasars, to
look for signs of scattered emission in the red wing: position angle rotation or
a change in the polarization level \citep{smi95,ogl97}. These data
could be used as a probe of the geometry in scenarios that the red
wing is scattered emission that was resonantly absorbed in an
out-flowing wind or it is scattered light from an optically thick
"moving mirror" that is created by a stream of clouds in motion.
\begin{acknowledgements}
I am indebted to Bev Wills for sharing the reduced HST data from
1992 and her ground based data from the same campaign. I would like
to thank Matt Malkan for sharing his expertise and encouraging me to
pursue this topic. Data from the
Steward Observatory spectropolarimetric monitoring project were
used. This program is supported by Fermi Guest Investigator grants
NNX08AW56G, NNX09AU10G, and NNX12AO93G. I was also very fortunate
to have a referee who greatly improved the clarity and focus of this effort.
\end{acknowledgements}

\end{document}